\documentclass[proceedings]{JHEP} 

\def\lsim{\mathrel{\rlap {\raise.5ex\hbox{$ < $}}
{\lower.5ex\hbox{$\sim$}}}}
\def\gsim{\mathrel{\rlap {\raise.5ex\hbox{$ > $}}
{\lower.5ex\hbox{$\sim$}}}}
\def\be{\begin{equation}}
\def\ee{\end{equation}}
\def\ba{\begin{eqnarray}}
\def\ea{\end{eqnarray}}
\def\bq{\begin{quote}}

\def\np#1#2#3{{\it {Nucl. Phys.}} {\bf{B#1}} (#2) #3}
\def\pl#1#2#3{{\it {Phys. Lett.}} {\bf{B#1}} (#2) #3}
\def\prl#1#2#3{{\it {Phys. Rev. Lett. }}{\bf{#1}} (#2) #3}
\def\pr#1#2#3{{\it {Phys. Rev.}} {\bf{D#1}} (#2) #3}

\conference{European Network on Physics beyond the Standard Model, 1999}

\title{Yukawa couplings and proton decay in SUSY models }

\author{{Mario  E.  G\'omez}\\

Physics Division, School of Technology,\\
Aristotle University of Thessaloniki, \\ 
Thessaloniki, GR-540 06, Greece\\
        E-mail: \email{mgomez@cc.uoi.gr}}

\abstract{We discuss proton decay induced by dimension--5 operators in 
supersymmetric models containing extra 
hypercharge--$1/3$ colour--triplets. We derive a 
general  formula relating dimension--5 operator to the colour--triplet mass matrix.
We show that certain zeros in the triplet mass--matrix together with 
some triplet coupling selection rules can lead to elimination of 
dimension--5 operators. We apply this mechanism to $SU(5)$ and flipped $SU(5)$ theories
with extended Higgs sectors.}
\begin{document}

\section{Introduction}

Proton decay is 
a generic feature of any unification scheme since the unification 
of quarks and 
leptons in a common multiplet introduces extra interactions 
that violate baryon number. 
Proton decay rates and modes are a prediction of GUT models that 
play a crucial role in 
their phenomenological viability. In fact, proton decay has 
turned out to be the 
nemesis of many GUT and Superstring models. It is a welcome prediction that can be used 
to test GUTs. In supersymmetric GUTs with conserved $R$--parity the dominant 
baryon number violating operators are dimension $D=5$, while $D=6$ operators 
are in general suppressed due 
to the increase of the unification scale in comparison to its 
non--supersymmetric values. $D=5$ 
 operators are proportional to the Yukawa couplings and to the inverse 
of the  heavy mass \cite{DIMF}. In minimal models the 
Yukawa couplings involved are associated with the fermion masses. The values 
of these couplings play an important role in the final value of the proton 
decay rate and the resulting hierarchy of existing modes. 
Nevertheless, Superstring embeddable models \cite{SEM} or 
models of phenomenologically oriented GUTs that treat the fermion mass problem
 \cite{FMM}, come out with an extended Higgs sector. 

In this talk  we summarize the results of a recent work	\cite{yo},
where  we propose a mechanism 
for eliminating or suppressing such operators based on the use of   
textures of the hypercharge 1/3 mass--matrix accompanied by certain 
constraints of the extra triplet coupling to matter.
 
\section{Proton Decay in minimal SU(5) models}
Let us consider unified models with the minimal Higgs content to allow the
beaking of the $SU(5)$ symmetry to  $SU(3)_c\times SU(2)_L\times U(1)_Y$ at 
the GUT scale $M_{GUT}$.

Non Supersymmetric $SU(5)$ models predict proton decay as a 
consequence of gauge 
interactions of the heavy particles, originated when the $SU(5)$ 
symmetry is broken,
and quarks and leptons. The baryon-number-vioalting operators are 
$D=6$ and they
are suppressed as the square of the mass of the heavy particles 
($\approx M_{GUT}$).
The dominant proton decay mode in these models is 
$p \rightarrow e^{+} \pi^0$. The calculated lifetime \cite{PN} is :
\begin{equation}
\tau(p\rightarrow e^+ \pi^0)\approx (\frac{M_{GUT}}
{3.5\cdot 10^4 {\rm GeV}})^4\times 10^{31\pm 1} yr
\end{equation}

While the experimental bound for this process is \cite{pdb}
\[
\tau(p \rightarrow e^{+} \pi^0) > 5.5 \times 10^{32} years
\] 

Since tipical values for the {\it cuasi} unification in non-susy 
$SU(3)_c\times SU(2)_L\times U(1)_Y$ 
are $M_{GUT}\approx 10^{13}-10^{14} Gev$, proton 
lifetime predictions exceeds the experimental bounds. 

In SUSY SU(5) models, the scale of unification is increased to 
$M_{GUT}\approx 10^{16} GeV$, 
this is enought to bring the prediction for $D=6$ mediated proton decay to a safe 
limit:
\[
\tau(p \rightarrow e^{+} \pi^0) \approx 10^{35\pm 1} years
\] 

However proton decay is predicted, at smaller rates, due to 
Yukawa interactions.  
In this case the supersymmetric partners of the colored triplet 
Higgs bosons interact
with leptons (sleptons) and quarks (squarks) fields. $D=5$ operators arises suppresed
only by one power of $M_{GUT}$.    

Color triplets are contained in  Higgs pentaplets
${h},{\overline{h}}$. The quarks and leptons are  assigned to 
$\phi({\bf\overline{5}})+\psi({\bf\overline{10}})$ representations of $SU(5)$.
The part of the superpotential related to dimension--5 decay will  be
\begin{equation}
Y^u_{ij} \,\psi_i \psi_j {h}_{1} + Y^d_{ij}\,\psi_i\phi_j
 {\overline{h}}_{1}+{\mu}{h}{\overline{h}}
+{\lambda}{h}{\Sigma}{\overline{h}}
\ ,
\label{abo}
\end{equation}
where the symbol  $\Sigma$ stands for 
the adjoint Higgs superfield in the {\bf{24}} representation.

The $SU(5)$ symmetry is broken down to the MSSM, when $\Sigma$ 
gets a VEV, V,  along the 
24--direction. The isodoublet and colour--triplet  masses are
\begin{equation}
{M}_2=\mu-3{\lambda}V \ ,
\end{equation}
\begin{equation}
{M}_3=\mu+2{\lambda}V
\end{equation}
 The triplets are heavy, ${M}_3\sim M_{GUT}$, while the 
doublet pair must remain light
${M}_2\sim m_w$. Hence a fine--tuning condition must be imposed 
in the parameters of the superpotential (\ref{abo}).

The effective $SU(3)_c\times SU(2)_L\times U(1)_Y$ superpotential 
describing the couplings of quarks and leptons to the extra 
coloured triplets of the $D$--quark type
\begin{equation}
Y^u_{ij}Q_i Q_j D+Y^d_{ij} Q_i L_j{\overline{D}} + Y^u_{ij}E_i^{c}U_j^{c}D \ ,
\end{equation}

$D=5$ operators can be converted into four--fermion operators by the 
appropiate gaugino dressing.  Assuming roughly an overall universal 
supersymmetry breaking scale  $m_S$ ,the corresponding four--fermion operator 
will involve:
\begin{equation}
\label{eq:op4}
\lambda \cdot\left[{\frac{({M}_3)^{-1}}{m_S}}
-4 m_S({{M}_3})^{-3}
\log{\frac{({{M}_{3}})}{m_S}}\right]
\end{equation} 

Where $\lambda$ contain a combination of Yukawa and gauge couplings.

The theoretical predictions \cite{PN} for the mode $p\rightarrow \overline{\nu} K$ 
are comparable to the experimental bound for this
mode \cite{pdb} 

\[
\tau(p\rightarrow \overline{\nu} K) > 5.5\times 10^{32} yr. 
\]

Hence, the parameter space for the minimal SUSY--$SU(5)$ is very
restricted.

In $SU(5)$--models with a non-minimal content of Higgs multiplets, 
${M}_3^{-1}$ of 
eq.~(\ref{eq:op4}) will be replaced by a matrix, and therefore 
its null elements will
play an important role in  the suppresion of 
$D=5$--operator mediated proton decay.

In the minimal flipped $SU(5)\times U(1)$ 
model \cite{BN} matter fields come in the representations 
\begin{equation}
{F}_i({\bf{10}}, 1/2)\,\,\,,\,\,\,\,\,{f}_i^c({\bf{\overline{5}}},-3/2)\,\,,
\,\,\,\,\,{l}_i^c({\bf{1}},5/2)
\end{equation}
while Higgses in 
\begin{equation}
{h}({\bf{5}},-1)\,\,\,,\,\,\,{\overline{h}}
({\overline{\bf{5}}}, 1)\ ,
\end{equation}
and in 
\begin{equation}
{F}_{h}({\bf{10}}, 1/2)\,\,\,,\,\,\,\,\,{\overline{F}}_h
({\overline{\bf{10}}},-1/2)\ .
\end{equation}

The part of the superpotential relevant for the beaking of the 
unifiying symmetry and Yukawa terms is:

\ba
&&f_{ij}F_i\,F_j\,{h}+y_{ij}F_i\,f_j^{c}
\overline{h}+r_{ij}l_i^{c}\,f_j^{c}\,{\overline{h}}+ \nonumber\\
&&\mu_{h} {\overline{h}}+
{\lambda}{F}_{h}{F}_{h}{h}
+{\overline{\lambda}}{\overline{F}}_{h}{\overline{F}}_{h}{\overline{h}}
\label{mfl}
\ea

VEV's of $F_h$ and $\overline{F_h}$ along the neutrino-like 
component break the 
$SU(5)\times U(1)$ symmetry to the MSSM.  A great advantage 
of the  ``flipped" $SU(5)$ model over the ordinary one is 
that of the realization of the ``triplet--doublet splitting" mechanism 
without fine--tuning the parameters of the superpotential (\ref{mfl}). In this 
case the doublet mass 
is given by the parameter $\mu$ which can be $\sim m_w$ while 
the mass matices for the triplets:
\begin{equation}
{\cal{M}}_3=\left(\begin{array}{c c}
0 & \lambda V \\
\overline{\lambda} \overline{V} & 0 \end{array}\right)
\end{equation}
Where the entry 22 is null since the pair $F_h\, {\overline{F}_h} $ 
has to be massless in order to realize the $SU(5)\times U(1)$ breaking 
to the standard model. Since in this model there is not $D \overline{D}$ mass term,
$D=5$-operators are naturally suppresed.

\section{How to suppress dimension--5 operators in effective models with extra triplets}

 Let us consider a general supersymmetric model containing some extra hypercharge--$1/3$ colour--triplets
\footnote{This superpotential arises in the case of the standard $SU(5)$ with extra
 Higgs 5--plets or from the 
flipped $SU(5)\times U(1)$ with both extra Higgs 5--plets and 10--plets.}.
The effective $SU(3)_c\times SU(2)_L\times U(1)_Y$ superpotential describing the 
couplings of quarks and leptons to the extra coloured triplets of the $D$--quark type
\begin{equation}
f_{ij}^{\alpha}Q_i Q_j D_{\alpha}+y_{ij}^{\alpha} Q_i L_j{\overline{D}}_{\alpha}
 + r_{ij}^{\alpha}E_i^{c}U_j^{c}{\overline{D}}_{\alpha} \ ,
\label{ww}
\end{equation}
where $i,j=1,2,3$ are the usual generation indices and $\alpha=1,...,N$ is an extra 
index describing the multiplicity of triplets
and repeated indices are summed. In addition the effective triplet mass matrix
will be of the form
\begin{equation}
{({\cal M}_3)}_{\alpha \beta}D_{\alpha}\overline{D}_{\beta}
\end{equation}
where ${\cal M}_3$ is in general non--diagonal.  
 
We can always go to a basis in which the triplet mass--matrix is diagonal
\begin{equation}
D_{\alpha}=S_{\alpha \beta}D_{\beta}^\prime\,\,\,\,,\,\,\,
\overline{D}_{\alpha} = U_{\alpha \beta}\overline{D}_{\beta}^\prime
\end{equation}
\begin{equation}
{{\cal M}_3}_D\equiv diag (m_1,m_2,\cdots,m_N)=S^{T}{{\cal M}_3}\,U
\end{equation}
where the matrices $S$ and $U$ are unitary.  In this basis we can easily evaluate
 $D=5$ operators resulting from Higgs triplet fermion exchange, and then recast the 
result in the original basis.
Assuming that all triplets are massive ($m_i\ne0\,,i=1,\dots, N$),
operators with the structure $Q_i\,Q_j\,Q_k\,L_n$ will be proportional to
\footnote{The corresponding four--fermion operator, assuming roughly 
an overall universal 
supersymmetry breaking scale  $m_S$ , will involve
 ${\frac{({\cal{M}}_3)^{-1}}{m_S}}
-4 m_S({{\cal{M}}_3})^{-3}
\log{\frac{({{\cal{M}}_{3}})}{m_S}}$ .} 
\ba
{\cal O}^{\mbox{\it\tiny QQQL}}_{ijkl}&=&
\sum_{\alpha,\beta,\gamma=1}^{N}f_{ij}^{\alpha}
S_{\alpha\gamma}({{\cal M}_3}_{D}^{-1})_{\gamma}U_{\beta\gamma}y_{kn}^{\beta}\nonumber\\
&=&\sum_{\alpha ,\beta =1}^{N}f_{ij}^{\alpha}({{\cal M}_3}^{-1})_{\alpha \beta}^{T}
y_{kn}^{\beta}\nonumber\\
&=&\frac{1}{\det({{\cal M}_3})}\,\sum_{\alpha,\beta=1}^{N}
f_{ij}^{\alpha}\,{\rm cof}({{\cal M}_3})_{\alpha\beta}\,y_{kn}^{\beta}
\label{ma}
\ea
Analogous formulas hold for $D=5$ operators of the type $Q_iQ_jU^c_kE^c_k$.

Suppose now that we want to eliminate all dimension five operators. 
Assuming that 
the Yukawa couplings $f_{ij}^\alpha$ and $y_{ij}^\beta$ are in general unrelated  and
$\det {\cal M}_3\ne0\,$, equation (\ref{ma}) implies that 
{\em 
the necessary and
sufficient condition for vanishing of the ${\cal O}^{\mbox{\it\tiny QQQL}}_{ijkl}$ operator is that
for every pair of triplets ($D^\alpha$,${\overline{D}}^\beta$\ ,
$\alpha,\beta=1,\dots,N$) that do couple to quarks and leptons 
 respectively ($f^\alpha_{ij}\ne0\ and\ h^\beta_{ij}\ne0$)
the cofactor of the corresponding triplet mass matrix element $({\cal M}_3)_{\alpha\beta}$
vanishes}\footnote{We consider here the triplet $D^\alpha$ as coupled 
to quarks and leptons if at least one 
 $f^\alpha_{ij}\ne0$ and similarly for anti--triplets.}
\ba
&&{\cal O}^{\mbox{\it\tiny QQQL}}_{ijkl}=0 \Longleftrightarrow
{\rm cof}({\cal M}_3)_{\alpha\beta}=0\nonumber\\&& \forall\ (\alpha,\beta)\in 
\Xi=\{(\alpha,\beta) : f^\alpha_{ij}\ne0\nonumber\\
&& {\rm and}\  h^\beta_{kl}\ne0\}\ .
\label{con}
\ea
It is obvious that in the case where all triplets ($D$'s and $\overline{D}$'s) 
couple to matter
the suppression of dimension five operators (\ref{ma}) is not possible since
(\ref{con}) leads to 
$\det(M_3)=0$.
Nevertheless, if for some reason (discrete symmetry, R--parity, anomalous $U(1)$, 
accidental symmetry) some of the $f^\alpha_{ij}$ and/or $y^\beta_{kl}$  are zero
and the triplet mass matrix is such that the cofactors of the appropriate 
 matrix elements are zero then the associated dimension--5 operator 
vanishes. 

The previous discussion leads to the possibility that {\em in a model with extra D--quark triplets
dimension--5 operators can be  eliminated by using  textures of triplet mass matrices
and the triplet--matter couplings.}

To be concrete let us give a simple example of such an effective theory. Consider 
the case of an effective theory with two extra triplets. Only the first 
couples  to matter through the superpotential terms
\begin{equation}
 f_{ij}^{1}Q_iQ_jD_{1}+y_{ij}^{1}Q_iL_j{\overline{D}}_{1}
+r_{ij}^{1}E_i^{c}U_j^{c}{\overline{D}}_{1}
\label{tsp}
\end{equation}
and their mass--matrix has the form
\be
{\cal M}_3=\left(\begin{array}{cc}\mu_{11}&\mu_{12}\\ \mu_{21}&0\end{array}\right)\ .
\label{tmm}
\ee
Since $f_{ij}^{2}=y_{ij}^{2}=0\,$ evaluation of (\ref{ma}) leads to
\be
{\cal O}^{\mbox{\it\tiny QQQL}}_{ijkl}= 
f_{ij}^{1}\,{\rm cof}({{\cal M}_3})_{11}\,y_{kn}^{1} \sim {\rm cof}({{\cal M}_3})_{11} =0
\ee
It is remarkable that if we remove the second pair of triplets of the model
(which do not couple directly to matter) the usual dimension--5 operators 
reappear.
We shall study below that this nice property can be incorporated 
 in $SU(5)$ models.

\section{$SU(5)$ models without dimension--5 operators}
 
Let us consider now an $SU(5)$ model with two pairs of Higgs pentaplets
${h}_\alpha,{\overline{h}}_\alpha\,, \alpha=1,2$ of which only the 
first couples to matter.
The quarks and leptons are  assigned to 
$\phi({\bf\overline{5}})+\psi({\bf\overline{10}})$ representations of $SU(5)$.
The 
part of the superpotential related to dimension--5 decay will  be
\ba
&&f_{ij} \,\psi_i \psi_j {h}_{1} + y_{ij}\,\psi_i\phi_j
 {\overline{h}}_{1}+\nonumber\\
&&\sum_{\alpha,\beta=1}^2({\mu}_{\alpha \beta}{h}_{\alpha}
{\overline{h}}_{\beta}
+{\lambda}_{\alpha \beta}{h}_{\alpha}{\Sigma}{\overline{h}}_{\beta})
\ ,
\label{su5s}
\ea
where the symbol  $\Sigma$ stands for 
the adjoint Higgs superfield in the {\bf{24}} representation.
The isodoublet and colour--triplet  mass matrices are correspondingly
 of the form
\begin{equation}
{\cal{M}}_2=\mu-3{\lambda}V \ ,
\end{equation}
\begin{equation}{\cal{M}}_3=\mu+2{\lambda}V\end{equation}
The well known fine--tuning that guarantees a massless pair of isodoublets amounts to 
\begin{equation}
\det({\cal{M}}_2)=0 \ .
\label{ft}
\end{equation}
The proton decay rate through $D=5$ operators is, according to equation (\ref{ma}),
 determined 
by the cofactor of the $1-1$ element  of the  triplet mass matrix 
\begin{equation}
{\rm cof}({\cal M}_3)_{11}={(\mu_{22}+2{\lambda}_{22}V)} \ .
\end{equation}
Hence, choosing $\mu_{22}=-2{\lambda}_{22}V$ dimension--5 operators vanish. 
This condition  is perfectly compatible with the previous fine--tuning condition (\ref{ft}). 
It is very interesting that 
proton decay through $D=5$ operators can be set to zero through a condition on the couplings 
\footnote{
Of course, proton decay still goes through at the (suppressed) rate of $D=6$ operators.}.

In the framework of our standard $SU(5)$ example the required zero in the 
inverse triplet mass 
matrix does not correspond to any symmetry and is in a sense a 
second fine--tuning.
 Nevertheless, the general conclusion is that zeros of the triplet mass matrix,
 perhaps attributable to symmetries, can stabilize the proton. 

The superpotential considered above in (\ref{su5s}) is not the most general one.
In fact,  the case that all 5--plets couple to matter cannot be reduced
to (\ref{su5s}) since it would require a different Higgs 5--plet rotation 
for each generation of matter. However, we should emphasize the fact that in $SU(5)$ models with non minimal 
Higgs content, the constraints imposed by proton decay on the parameter space 
and triplet masses can be relaxed.


\section{\label{secf}Dimension--5 operators in extensions of the flipped $SU(5)$ }
 
In spite of the nice features of the minimal flipped $SU(5)$ model, 
all attempts to obtain such a model from strings have yielded up to now
non--minimal models. Such models include\\
(a) extra pairs of low energy Higges ($h$, $\bar h$) and/or\\
(b) extra pairs of $SU(5)\times U(1)$ breaking Higges ($F_h$, ${\overline{F}_h}$).

We are thus motivated to study the presence of dimension--5 operators in
such models. 
As we shall see contrary to the minimal case, such extensions
of the flipped $SU(5)$ model are not automatically free of dimension--5 
operators. 

The relevant part of the superpotential assuming $N_5$ pairs of Higgs 
5--plets 
($h_\alpha, {\overline{h}_\alpha}\ , \alpha=1,\dots,N_5$) that couple to matter and
$N_{10}$ pairs of Higgs 10--plets
($F_{h\alpha}, {\overline{F}_{hA}}\ , A=1,\dots,N_{10}$) that do not couple to matter, will
have the form
\begin{eqnarray}
&&
f_{ij}^{\alpha}F_i\,F_j\,{h}_{\alpha}+y_{ij}^{\alpha}F_i\,f_j^{c}
{\overline{h}}_{\alpha}+\nonumber\\
&&r_{ij}^{\alpha}l_i^{c}\,f_j^{c}\,{\overline{h}}_{\alpha}+
{\mu_{\alpha\beta}}{h}_{\alpha} {\overline{h}}_{\beta}+
m_{AB}{F}_{hA}{\overline{F}}_{hB}\nonumber\\
&&
\mbox{}+{\lambda}_{AB\gamma}
{F}_{hA}{F}_{hB}{h}_{\gamma}
+{\overline{\lambda}}_{AB\gamma}
{\overline{F}}_{hA}{\overline{F}}_{hB}{\overline{h}}_{\gamma}
\end{eqnarray}
where $A, B=1,\cdots,N_{10}$ ,  $\alpha, \beta, \gamma=1,\cdots,N_5\,$.
Assuming GUT symmetry breaking to an arbitrary direction in the Higgs 
$10$--plet space \newline
($(V_1,V_2,\dots,V_{N_{10}})$ and similarly for bars)
\footnote{D--flatness requires $\sum_A V_A^2=\sum_A{\overline{V}_A}^2$}, 
we obtain the triplet mass matrix
\footnote{In a $(D_1,\cdots,D_{N_5}, (d_H^c)_1,\cdots,(d_H^c)_{N_{10}})$ 
versus $({\overline{D}}_1,\cdots,
{\overline{D}}_{N_5}, ({\overline{d}}_H^c)_1,\cdots, 
({\overline{d}}_H^c)_{N_{10}})$ basis, where with  $D$ we denote
the triplets which lie inside the Higgs 5--plets and with $D_{H}$ the triplets
that lie inside the Higgs 10--plets. }
\begin{equation} 
{\cal{M}}_3=\left(\begin{array}{cc}
{\mu}_{\alpha\beta}&v_{\alpha A}\\
{\overline{v}}_{A\beta}&m_{AB}
\end{array}\right)
\end{equation}
where $\mu_{\alpha\beta}$ is the doublet mass--matrix and
 $v_{\alpha A}=2{\lambda_{AB\alpha}} V_B\,$,
${\overline{v}}_{A\beta}=2{\overline{\lambda}}_{AB\beta} {\overline{V}}_B\,$

F--flatness demands ${\det(m)=0}$ in order to have at least one pair of massless 
Higgs decuplets which will realize the GUT symmetry breaking.
One can actually choose $m$ to have only one zero eigenvalue so that 
all remnants of the Higgs decuplets will become heavy.

Let us now start our study by a simple example.
Consider the flipped model with two pairs of Higgs 5--plets and one pair of 
Higgs 10--plets.
Assuming for simplicity that the 5-plet mass matrix is diagonal,
the explicit form of the triplet matrix is 
\footnote{We have renamed $\lambda_1=\lambda_{111}\,, \lambda_2=\lambda_{112}$.}
\begin{equation}
{\cal{M}}_3=\left(\begin{array}{ccc}
0&0&{\lambda}_1V\\
0&\mu&{\lambda}_2V\\
{\overline{\lambda}}_1{\overline{V}}&{\overline{\lambda}}_2{\overline{V}}&0
\end{array}\right)
\end{equation}
and $\det({\cal M}_3)=\lambda_1{\overline{\lambda}}_1{\overline{V}}$
The transpose of inverse triplet matrix entering in formula (\ref{ma}) is 
\begin{equation}
\left({{\cal{M}}_3^{-1}}\right)^T= \left(
\begin{array}{ccc}
{\frac{\lambda_2{\overline{\lambda}}_2}{\lambda_1{\overline{\lambda}}_1\mu}}&
-\frac{{{\lambda}}_2} {{{\lambda}}_1\mu}&
\cdot\\
-\frac{{\overline{\lambda}}_2}{{\overline{\lambda}}_1\mu}
&\frac{1}{\mu}&\cdot\\
\cdot&\cdot&\cdot
\end{array}\right)
\label{myma}
\end{equation}
where the dots  stand for elements which are irrelevant. It is now obvious that in
this  model dimension five operators cannot be eliminated since even in the
case $\overline{\lambda}_2=\lambda_2=0$ the $22$ element does not vanish.
If we want to eliminate them we have two solutions :\\
(i) assume that the extra pair of 5--plets does not couple to matter. In this
case only the $11$ element of the matrix in (\ref{myma}) is relevant and 
it vanishes for $\lambda_2=0$ (or ${\overline{\lambda}_2=0}$).\\
(ii) make the milder assumption that one of the 5--plets (e.g. $h_2$)
 does not couple to the up quarks (or similarly $\overline{h}_2$ does not couple
to the down).  In this case the second column (or line) of the matrix in 
(\ref{myma}) becomes irrelevant and  the column (or line) left vanishes for
 $\lambda_2=0$ (or $\overline{\lambda}_2=0$).

Another case that could arise is the existence of extra decuplets. The simplest of 
these cases is for $N_5=1$ and $N_{10}=n\geq 2$. 

\be
{\rm cof}({\cal M}_3)_{11}=\frac{\det m}{\det{\cal M}_3}
\ee
which means that proton decay is absent only in the case that the restricted
  mass--matrix of the triplets not coupled to matter has 
\be
\det m =0
\label{cc}
\ee
This constrain naturally arises in the
context of the flipped $SU(5)\times U(1)$ model as consequence of 
F--flatness  as we mentoned above.

In the more general case where $N_5$ and $N_{10}$ are arbitrary 
dimension--5 operators can be suppressed only in the case $N_5\le N_{10}$ . Furthermore 
one has to require 
that the Higgs decuplet mass matrix has $N_5$ zero eigenvalues.
This is compatible with symmetry breaking and with the requirement of 
making all triplets heavy but leaves $N_5-1$ pairs of $Q({\bf3},{\bf2},1/6)+
{\overline Q}({\bf\bar3},{\bf2},-1/6)$ massless.
This feature does not necessarily mean that this possibility is ruled out.
On the contrary one can consider the cases where extra $Q$'s have intermediate 
masses which are small enough to sufficiently suppress dimension--5 operators
but they are still compatible with renormalization group requirements.
The appearance of extra vector--like pairs of $Q$ and $D$ type multiplets with intermediate 
masses is a welcomed feature in the context of flipped $SU(5)\times U(1)\,$ 
models that raise the unification scale to the string scale \cite{LN}. 

\section{Conclusions }

Our main result is that textured zeros of the color--triplet mass--matrix as well as 
 Yukawa selection rules can eliminate certain dimension--5 operators. In order to be specific we focused on $SU(5)$ 
 models. 
 In particular, we showed that introducing an extra pair of Higgs pentaplets in the standard supersymmetric $SU(5)\,$,
  with specific couplings, can eliminate 
 these operators. We also considered the case of the flipped--$SU(5)$ model with extra pentaplets and decuplets 
 and analyzed the conditions for vanishing proton decay through dimension--5 operators. Flipped--$SU(5)$ with 
 extra decuplets was shown to be $D=5$ operator--free as it happens in the
case of the minimal model.
However, flipped--$SU(5)$ with extra Higgs pentaplets is
not automatically free of dimension--5 operators. 
We have proposed a solution to this 
problem which  involves a texture of the pentaplet matrix together 
with certain constraints on the pentaplet couplings to matter.

{\bf{Acknowledgments}}

This work is supported by E.U. under the TMR contract 
``Beyond the Standard Model'', 
ERBFMRX-CT96-0090.


\begin{thebibliography}{99}


\bibitem{DIMF} S. Weinberg, \pr{26}{1982}{287};\\
N. Sakai and T. Yanakida, \np{197}{1982}{533}.

\bibitem{SEM} see e.g.\\
I. Antoniadis and G. Leontaris, \pl{216}{1989}{333};\\
S. Ranfone and J. W. F. Valle,
\pl{386}{1996}{151};\\
J. Rizos and K. Tamvakis, \pl{414}{1997}{277}.

\bibitem{FMM} see e.g.
K. S. Babu and S. M. Barr, \prl{75}{1995}{2088}.

\bibitem{yo} M. E. G\'omez, J. Rizos and K. Tamvakis
 {\it Phys. Rev. }{\bf D59}
(1998)015015.

\bibitem{PN} P. Nath and  R. Arnowitt, 
hep-ph/9808465 and references therein.

\bibitem{pdb}
C. Caso {it et al.}, The European Physics Journal {\bf C3}(1998)1.


\bibitem{BN} S. M. Barr, \pl{112}{1982}{219};\\
 J. P. Derendinger, J. H. Kim, D. V. Nanopoulos, \pl{194}{1987}{231}.

\bibitem{LN} J. L. Lopez and D. V. Nanopoulos, \np{399}{654}{1993};\\
see also J. L. Lopez, D. V. Nanopoulos and A. Zichichi, hep-ph 9307211 and
references therein.

\end{thebibliography}
 \end{document}